\documentclass[preprint,12pt]{elsarticle}

\usepackage{algorithm}
\usepackage{algpseudocode}
\usepackage{amssymb}
\usepackage{amsmath}
\usepackage{cleveref}
\usepackage{multirow}
\usepackage{subcaption}
\usepackage[table]{xcolor}
\usepackage{xspace}

\journal{Computers in Biology and Medicine}

\newcommand{\ours}{AFTNet\xspace}
\begin{document}

\begin{frontmatter}

\title{Deep Learning-based MRI Reconstruction with Artificial Fourier Transform Network (\ours)}
\author[1]{Yanting Yang}
\author[1]{Yiren Zhang}
\author[1]{Zongyu Li}
\author[2]{Jeffery Siyuan Tian}
\author[1]{Matthieu Dagommer}
\author[3]{Jia Guo}
\ead{jg3400@columbia.edu}
\affiliation[1]{
    organization={Department of Biomedical Engineering, Columbia University},
    addressline={500 W. 120th Street \#351},
    city={New York},
    postcode={10027},
    state={NY},
    country={United States}
}
\affiliation[2]{
    organization={Department of Computer Science, University of Maryland},
    addressline={8125 Paint Branch Drive},
    city={College Park},
    postcode={20742},
    state={MD},
    country={United States}
}
\affiliation[3]{
    organization={Department of Psychiatry, Columbia University},
    addressline={1051 Riverside Drive},
    city={New York},
    postcode={10032},
    state={NY},
    country={United States}
}

\begin{abstract}
Deep complex-valued neural networks (CVNNs) provide a powerful way to leverage complex number operations and representations and have succeeded in several phase-based applications. However, previous networks have not fully explored the impact of complex-valued networks in the frequency domain. Here, we introduce a unified complex-valued deep learning framework—Artificial Fourier Transform Network (\ours)—which combines domain-manifold learning and CVNNs. \ours can be readily used to solve image inverse problems in domain transformation, especially for accelerated magnetic resonance imaging (MRI) reconstruction and other applications. While conventional methods typically utilize magnitude images or treat the real and imaginary components of k-space data as separate channels, our approach directly processes raw k-space data in the frequency domain, utilizing complex-valued operations. This allows for a mapping between the frequency (k-space) and image domain to be determined through cross-domain learning. We show that \ours achieves superior accelerated MRI reconstruction compared to existing approaches. Furthermore, our approach can be applied to various tasks, such as denoised magnetic resonance spectroscopy (MRS) reconstruction and datasets with various contrasts. The \ours presented here is a valuable preprocessing component for different preclinical studies and provides an innovative alternative for solving inverse problems in imaging and spectroscopy. The code is available at: https://github.com/yanting-yang/AFT-Net.
\end{abstract}


\end{frontmatter}

\section{Introduction}

The shift from real to complex coordinate space in deep neural networks has unveiled the potential of complex numbers’ rich representational capacity, thereby spurring the development of complex-valued neural architectures~\cite{georgiou1992complex,guberman2016complex,trabelsi2018deep}. A corresponding, though inverse, domain shift occurs in the preprocessing of magnetic resonance imaging (MRI), where raw data are acquired and stored in complex-valued k-space, with each pixel encoding spatial frequency information across two or three dimensions. Following acquisition, the raw k-space data are transformed into images via a reconstruction process that enables interpretation by MR operators, physicians, radiologists, or data scientists. This crucial reconstruction step underpins overall image quality. Proper image reconstruction methods can increase the signal-to-noise ratio (SNR) by suppressing thermal noise~\cite{johnson1928thermal,nyquist1928thermal}, improve spatial inhomogeneities affected by point spread functions (PSFs), and correct unexpected signal artifacts~\cite{hansen2015image}.

Theoretically, image reconstruction is performed using domain transforms, for example Fourier transforms for fully sampled Cartesian data~\cite{fessler2010model}. However, in clinical settings where signal non-idealities are prevalent, numerical methods and machine learning approaches become indispensable. Traditionally, human experts select task-related features and develop models to capture the mapping between k-space and image domains~\cite{de2016machine}. However, due to significant pathological variations and the risk of human oversight~\cite{shen2017deep}, a consistent and unbiased diagnosis cannot be guaranteed. In recent years, k-space, as a low-dimensional feature space, has been leveraged in deep neural networks to learn the manifold mapping of domain transforms in low signal-to-noise settings~\cite{zhu2018image}. This image reconstruction process can be reformulated as a data-driven supervised learning task that determines the mapping between the k-space and the image domains, demonstrating superior immunity to noise and reconstruction artifacts. The conventional discrete Fourier transform algorithm, which is mathematically derived and is not based on learning, can be substituted by neural networks~\cite{lopez2021spiking}, as neural networks can learn complex mappings and priors from data, enabling them to recover missing information and reduce artifacts more effectively. This is particularly important in highly undersampled scenarios, where traditional DFT-based methods struggle to achieve high-quality reconstructions. Thus, fundamental neural networks have been presented that avoid the difficulties of finding the network structure and optimizing the algorithm. A similar approach is adopted in domain transform manifold learning along the phase-encoding direction~\cite{eo2020accelerating}, where the front-end convolutional layers, an intermediate global transform, and the back-end convolutional layers are combined to perform data restoration in the k-space and image domains. Modern score-based diffusion models provide a powerful way to sample data from a conditional distribution given the measurements in the k-space domain, which can be used to solve inverse problems in imaging~\cite{chung2022score}.

Although most deep learning-based MR image reconstruction algorithms have applied the concept of domain-transform learning that directly learns a low-dimensional joint manifold between the k-space and image domains, including end-to-end~\cite{zhao2023swingan} and the unrolled~\cite{yan2023dc} network, few works have fully leveraged complex-valued neural networks (CVNNs), which allow neural networks to implement learning-based frequency selection~\cite{xu2020learning}. Some early works proposed CVNNs but focused mainly on solving the basics of learning~\cite{hirose2012complex,hirose2012generalization}. In recent years, extensive studies have been conducted on complex-valued CNNs. Generalizations of real-valued CNN models have been shown to be significantly less vulnerable to overfitting~\cite{guberman2016complex}. Mathematical arguments and implementations have also been discussed~\cite{tygert2016mathematical,trabelsi2018deep}, enabling the practical application of CVNNs. A combination of CVNNs and vision deep learning models (e.g., UNet~\cite{ronneberger2015u}) has recently emerged and is being exploited for MR image reconstruction~\cite{sikka2021cu,cole2021analysis}, demonstrating superior and accelerated reconstruction compared to real-valued neural networks. However, the main drawback of previous CVNN works is that the potential of leveraging CVNNs in domain manifold learning has not been fully investigated.

\begin{figure}[t]
\centering
\includegraphics[width=\linewidth]{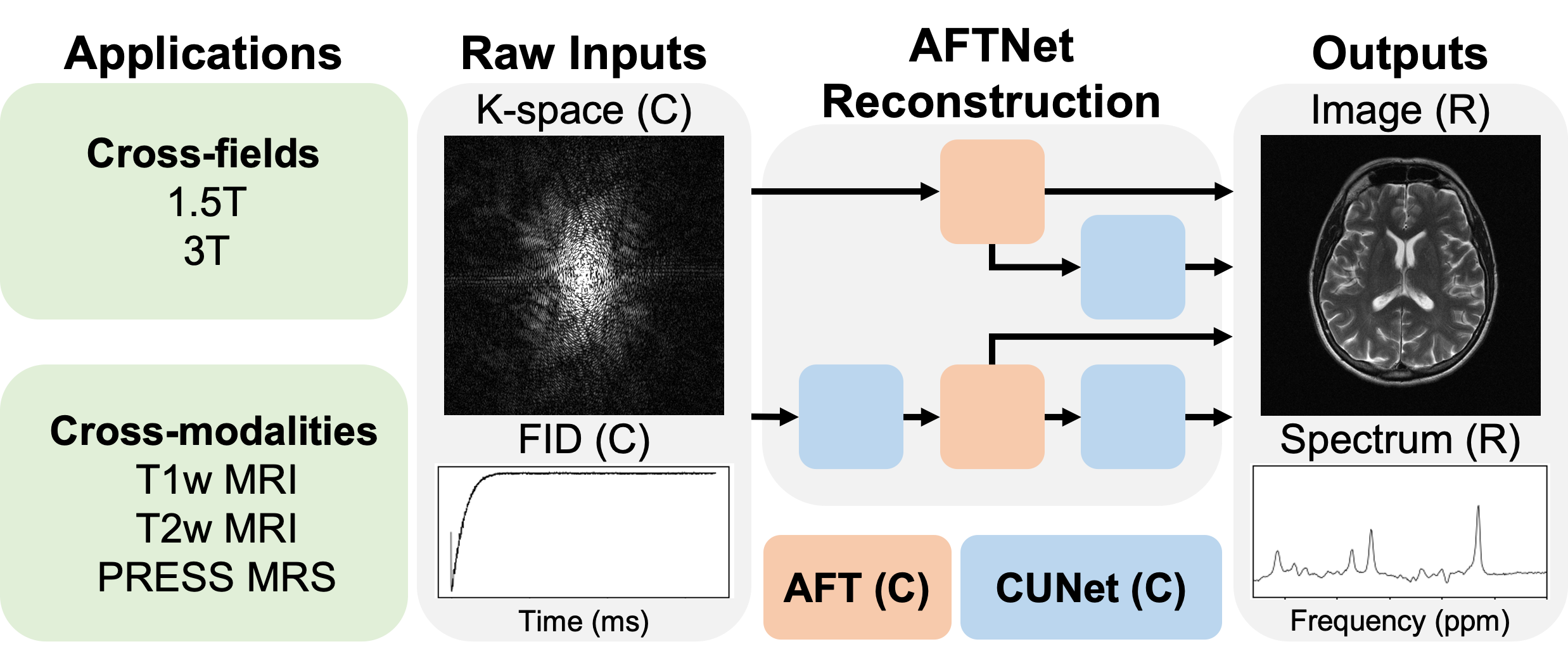}
\caption{Schematics of general deep-learning MR imaging/spectroscopy reconstruction based on \ours. The inputs of \ours can be 2D MRI k-space data or 1D MRS FID data. The outputs are reconstructed MR images or spectra. Different structures of \ours are developed by appending front-end and/or back-end convolutional networks to the AFT block. Here we show the T2w 1.5T MRI and 3T PRESS MRS reconstruction results, respectively. C: complex-valued and R: real-valued.}
\label{fig:main}
\end{figure}

To fully exploit the rich information contained in complex-valued MR data and overcome the limitations of traditional reconstruction methods, we propose (1) Artificial Fourier Transform (AFT) block which replaces conventional DFT with a learnable module and can adapt to noise and signal non-idealities, allowing the network to learn complex mappings between the k-space and image domains; (2) Artificial Fourier Transform Network (\ours), shown in \Cref{fig:main}, which combines domain-manifold learning with CVNNs by integrating AFT with complex-valued convolutional encoder-decoder networks to facilitate MR reconstruction. \ours directly process the full complex information (both magnitude and phase) inherent in k-space, avoiding the common pitfall of separating or discarding phase information. By operating directly on raw k-space data, \ours minimizes information loss and learns subtle frequency domain features critical for accurate image reconstruction. The CVNNs in \ours process global features in the frequency (k-space) domain and local details in the image domain, leading to improved artifact removal and reconstruction quality, particularly in highly undersampled scenarios. Through its modular design, \ours can be extended to any dimension (e.g., 1D MR spectroscopy data). We conduct extensive experiments in various modalities, system field strengths, acceleration ratios, and noise levels on the MRI reconstruction, accelerated reconstruction and MRS denoised reconstruction, achieving superior performance in all scenarios.

\section{Background}

\subsection{Complex-valued neural networks}\label{sec:cnet}

The definition of a conventional real-valued neural network can be extended to the complex domain. We denote a complex operator as $\mathrm{W}(\cdot) = \mathrm{W}_{real}(\cdot) + i\mathrm{W}_{imag}(\cdot)$, where $\mathrm{W}_{real}$ and $\mathrm{W}_{imag}$ are real-valued operators. The input complex vector can be represented as $z = \Re(z) + i\Im(z)$. The output of complex operator $\mathrm{W}$ acting on $z$ is derived following the vector dot product:
\begin{equation}\label{eq:basic}
\begin{split}
  y & = \mathrm{W} \cdot z \\
    & = \left[ \mathrm{W}_{\text{real}}(\Re(z)) - \mathrm{W}_{\text{imag}}(\Im(z)) \right] \\
    & \quad + i \left[ \mathrm{W}_{\text{imag}}(\Re(z)) + \mathrm{W}_{\text{real}}(\Im(z)) \right].
\end{split}
\end{equation}

As the linear operator and convolution operator are distributive~\cite{trabelsi2018deep}, we can directly replace $\mathrm{W}(\cdot)$ with $\mathrm{Linear}(\cdot)$ or $\mathrm{Conv}(\cdot)$ to obtain the complex-valued version:
\begin{equation}
\begin{split}
  \mathbb{C}\mathrm{Linear}(z) & = \left[\mathrm{Linear}_1(\Re(z)) - \mathrm{Linear}_2(\Im(z))\right] \\
  & \quad + i \left[\mathrm{Linear}_2(\Re(z)) + \mathrm{Linear}_1(\Im(z))\right]
\end{split}
\end{equation}
and
\begin{equation}\label{eq:conv}
\begin{split}
  \mathbb{C}\mathrm{Conv}(z) & = \left[\mathrm{Conv}_1(\Re(z)) - \mathrm{Conv}_2(\Im(z))\right] \\
  & \quad + i\left[\mathrm{Conv}_2(\Re(z)) + \mathrm{Conv}_1(\Im(z))\right]
\end{split}
\end{equation}
where we use subscripts $1$ and $2$ instead of $real$ and $imag$ to avoid misleading.

The complex version of the ReLU activation function we used in this study simply applies separate ReLU on both the real and the imaginary part of the input as follows:
\begin{equation}
\begin{split}
  \mathbb{C}\mathrm{ReLU}(z) & = \mathrm{ReLU}(\Re(z))+i\mathrm{ReLU}(\Im(z)))
\end{split}
\end{equation}
which satisfies Cauchy–Riemann equations~\cite{trabelsi2018deep} when both the real and the imaginary parts have the same sign or $\theta_z \in [0,\frac{1}{2}\pi] \cup [\pi,\frac{3}{2}\pi]$.

\begin{algorithm}[t]
\caption{Complex group normalization}\label{alg:gn}
\begin{algorithmic}
\State \textbf{Input}: $z$, $\gamma$, $\beta$, $G$ (number of groups), $\epsilon$ 
\State Shape of $z$ is $(B,C,H,W)$
\State Reshape $z$ to $(B,G,C//G,H,W)$
\For{each group}
    \State $\tilde{z} \gets V^{-\frac{1}{2}}(z-\mathbb{E}(z))$ based on \Cref{eq:gn2,eq:gn3,eq:gn4}.
\EndFor
\State Reshape $\tilde{z}$ to $(B,C,H,W)$
\State \Return $\gamma\tilde{z}+\beta$
\end{algorithmic}
\end{algorithm}

Normalization is a common technique widely used in deep learning to accelerate training and reduce the statistical covariance shift~\cite{ioffe2015batch,wu2018group,ba2016layer}. This is also mirrored in the CVNNs, where we want to ensure that both the real and the imaginary parts have equal variance. Extending the normalization equation to matrix notation we have:
\begin{equation}\label{eq:gn1}
  \tilde{z} = V^{-\frac{1}{2}}(z-\mathbb{E}(z))
\end{equation}
where $x-\mathbb{E}(x)$ simply zero centers the real and the imaginary parts separately
\begin{equation}\label{eq:gn2}
  z-\mathbb{E}(z) = \begin{bmatrix} \Re(z) - \mathrm{Mean}(\Re(z)) \\ \Im(z) - \mathrm{Mean}(\Im(z)) \end{bmatrix}
\end{equation}
and $V$ is the covariance matrix of real and imaginary parts of $z$
\begin{equation}\label{eq:gn3}
\begin{split}
  V &= \begin{bmatrix} V_{rr} & V_{ri} \\ V_{ir} & V_{ii} \end{bmatrix} +\epsilon I \\
  &= \begin{bmatrix} \mathrm{Cov}(\Re(z),\Re(z)) & \mathrm{Cov}(\Re(z),\Im(z)) \\ \mathrm{Cov}(\Im(z),\Re(z)) & \mathrm{Cov}(\Im(z),\Im(z)) \end{bmatrix}+\epsilon I.
\end{split}
\end{equation}

$V$ is a $2\times2$ matrix, and the existence of the inverse square root is guaranteed by adding $\epsilon I$ (Tikhonov regularization). Therefore, the solution of the inverse square root can be expressed analytically as
\begin{equation}\label{eq:gn4}
V = \begin{bmatrix} A & B \\ C & D \end{bmatrix} \Rightarrow V^{-\frac{1}{2}} = \begin{bmatrix} (D+s)/d & -B/d \\ -C/d & (A+s)/d \end{bmatrix}
\end{equation}
where $s=\sqrt{AD-BC}$, $t=\sqrt{A+D+2s}$ and $d=st$.

The complex normalization is defined as
\begin{equation}
\mathrm{Norm}(z) = \gamma\tilde{z}+\beta = \begin{bmatrix} \gamma_{rr} & \gamma_{ri} \\ \gamma_{ri} & \gamma_{ii} \end{bmatrix}\tilde{z}+\begin{bmatrix} \Re(\beta) \\ \Im(\beta) \end{bmatrix}.
\end{equation}
where $\gamma$ and $\beta$ are learnable parameters.

Considering the limitation of GPU RAM and the large memory consumption of complex-valued networks, we use group normalization~\cite{wu2018group} in our framework to avoid the possible inaccurate estimation of batch statistics caused by a small batch size. The group normalization in the complex domain can be represented as \Cref{alg:gn}.

\section{Methods}

\subsection{Artificial Fourier Transform}\label{sec:AFT}

\begin{figure}[t]
\centering
\includegraphics[width=\linewidth]{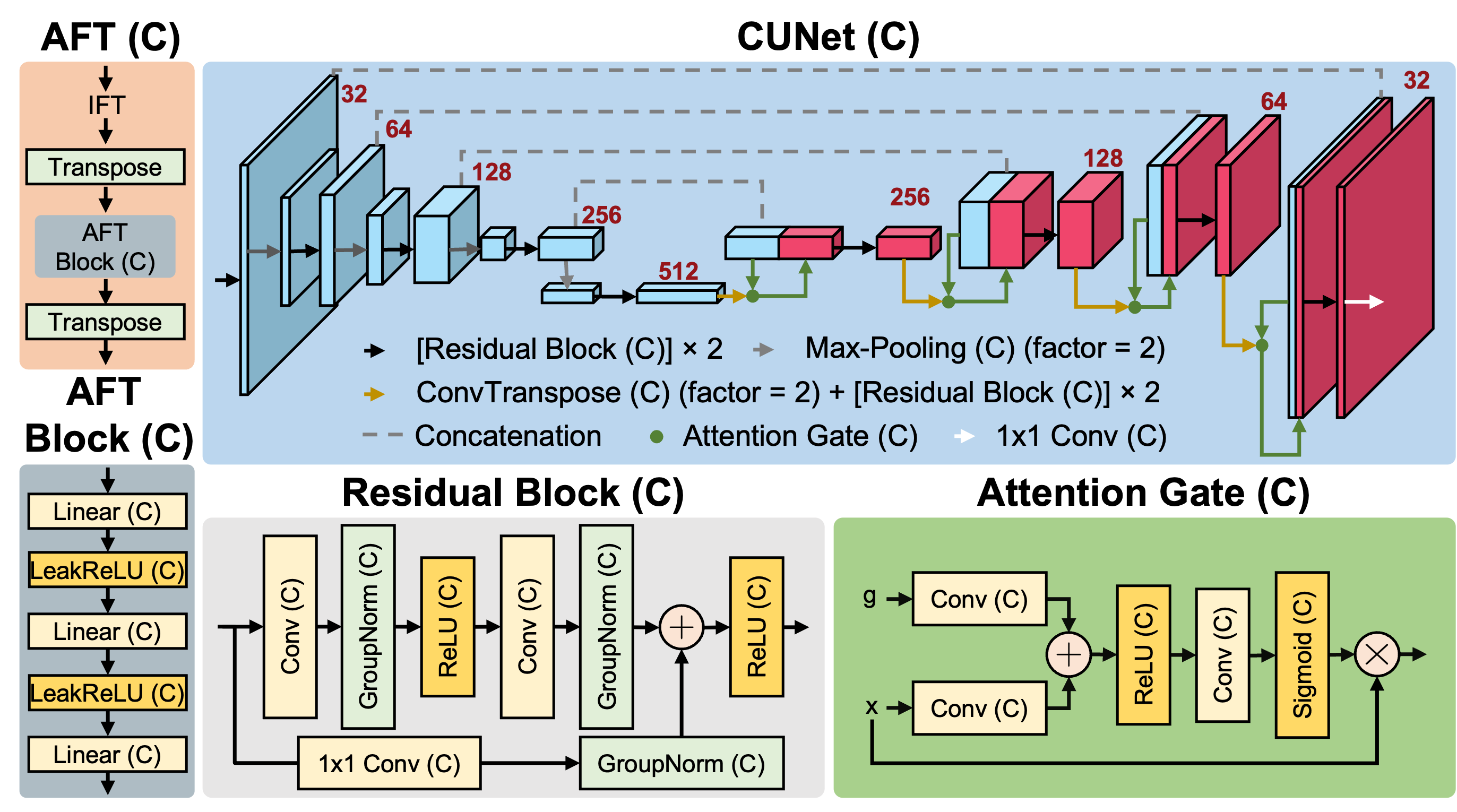}
\caption{Structure of 2-dimensional \ours. Components include the complex-valued AFT block, the complex-valued residual attention UNet, the complex-valued residual block, and the complex-valued attention gate. All convolutional layers have a kernel size of 3, except those pointed out specifically. C: complex-valued. Red numbers indicate the number of channels produced by each layer.}
\label{fig:architecture}
\end{figure}

Take cartesian sampling which is widely used in conventional MRI acquisitions as an example, since 2D discrete Fourier transform (DFT) is a linear operation and can be represented by two successive 1D DFTs as 
\begin{equation}
\begin{split}
\mathcal{F}_{x,y}\{f(x,y)\} &= \mathcal{F}_x\{\mathcal{F}_y\{f(x,y)\}\} \\
&= \mathcal{F}_y\{\mathcal{F}_x\{f(x,y)\}\},
\end{split}
\end{equation}
where the $x$ and $y$ are two separable independent variables. Each dimension of 2D DFT can be modeled as a trainable neural network~\cite{lopez2021spiking}. We further proposed that this idea could be naturally implemented with CVNNs shown in \Cref{fig:architecture}.

From the definition of the discrete Fourier transform of a sequence of $N$ complex numbers which can be represented in the real and imaginary parts as
\begin{equation}\label{eq:dft}
Z_k = \sum_{n=0}^{N-1} z_n \left[\cos\left(\frac{2 \pi}{N}kn\right) - i \sin\left(\frac{2 \pi}{N}kn\right)\right],
\end{equation}
rewrite \Cref{eq:dft} as
\begin{equation}
Z_k = W_{real}z + i W_{imag}z,
\end{equation}
where $W_{real}$ and $W_{imag}$ are the real and the imaginary coefficients. Use matrix notation to represent the real and imaginary parts of the DFT operation. We have the following:
\begin{equation}\label{eq:dft2}
\begin{bmatrix} \Re(Z_k) \\ \Im(Z_k) \end{bmatrix} = \begin{bmatrix} W_{real} & -W_{imag} \\ W_{imag} & W_{real} \end{bmatrix}\begin{bmatrix} \Re(z) \\ \Im(z) \end{bmatrix}.
\end{equation}

Compared \Cref{eq:dft2} with \Cref{eq:basic}, a multilayer CVNN with linear activation function can represent 1D DFT with appropriate weights. We use $AFT_N$ to denote the complex-valued Fourier transform deep learning block on the input vector with $N$ elements. The Fourier transform of the input data with dimension $H \times W$ can be represented as
\begin{equation}
Z = \mathrm{AFT}_H(\mathrm{AFT}_W(z)^\intercal)^\intercal.
\end{equation}

\subsection{Network framework}\label{sec:framework}

\begin{figure}[t]
\centering
\includegraphics[width=\linewidth]{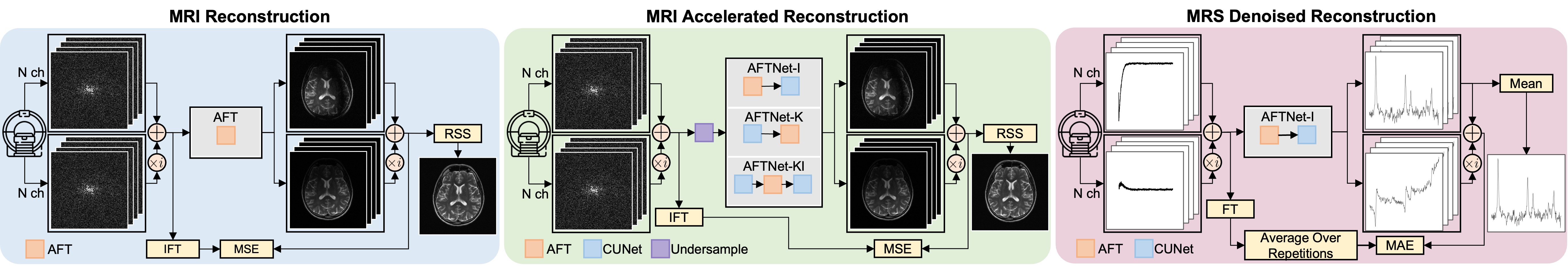}
\caption{The workflows of experiments on each dataset.}
\label{fig:workflow}
\end{figure}

The network structure and general workflow are shown in \Cref{fig:architecture} and \Cref{fig:workflow}. The AFT block is composed of an MLP with three complex linear layers linked by two complex LeakReLU activation layers. As for conventional accelerated MRI acquisition, the undersampling is applied to the phase-encoding direction, and we also only apply the AFT block to that direction. For the accelerated reconstruction task, we combine our AFT with an entirely complex-valued UNet (CUNet)~\cite{sikka2021cu} with residual attention, which extracts local and global features in the k-space and/or image domain. Multiple network architectures are evaluated to verify the effectiveness of both AFT and CUNet in different domains. We refer each of them to \textbf{AFT}, \textbf{\ours-I}, \textbf{\ours-K}, and \textbf{\ours-KI}, respectively as shown in \Cref{fig:workflow}.

The architecture of the CUNet presented in \Cref{fig:architecture} is generally based on the residual attention UNet~\cite{ronneberger2015u} but with all the components of real value replaced by complex value components as shown in \Cref{fig:architecture}, including complex-valued convolutional layers and complex-valued ReLU layers introduced in \Cref{sec:cnet}. We further optimize the network for smaller batch sizes by replacing batch normalization with group normalization as illustrated in \Cref{alg:gn}. Other complex components are implemented in the same way. For example, the complex transposed convolution operator can be mirrored from \Cref{eq:conv}, the complex sigmoid is applied like the complex ReLU, and the complex max pooling is almost the same as the real-valued version except that the indices are inferred from absolute values.

\subsection{Reconstruction Loss}

In the context of image reconstruction and processing, the impact of the loss function is vital if human observers are to evaluate the final results. One common and safe choice is $\ell_2$ loss which works under the assumption of Gaussian white noise. For training AFT for MRI reconstruction, the loss value is determined in the frequency domain as 
\begin{equation}
\mathcal{L}^{recon} = \mathcal{L}^{\ell_2}(\Re(x),\Re(y))+\mathcal{L}^{\ell_2}(\Im(x),\Im(y))
\end{equation}
so that both real and imaginary outputs are optimized to match the conventional Fourier transformation. $x$ and $y$ are predicted and targeted complex-valued images. For training \ours for accelerated MRI reconstruction, we also want to minimize the error of magnitude images. Therefore, the loss value for accelerated MRI reconstruction is
\begin{equation}
\mathcal{L}^{acc} = \mathcal{L}^{recon}+\mathcal{L}^{\ell_2}(\operatorname{RSS}(x),\operatorname{RSS}(y))
\end{equation}
where the root-sum-of-squares (RSS) approach~\cite{roemer1990nmr} is applied to complex-valued output from the model to generate the optimal, unbiased estimate of magnitude image which is used for loss calculation. For denoised MRS reconstruction, we use $\ell_1$ loss following the previous practice~\cite{ma2024magnetic} so the loss is
\begin{equation}
\mathcal{L}^{denoise} = \mathcal{L}^{\ell_1}(\Re(x),\Re(y))+\mathcal{L}^{\ell_1}(\Im(x),\Im(y))
\end{equation}

Data consistency is applied after each RACUNet and AFT block so that the known positions of the output data are replaced by the original data samples to obtain better fidelity~\cite{wu2023deep}.

\section{Experiments}

\subsection{Overview}

In the MRI reconstruction experiment, we apply our AFT to the multicoil k-space data acquired directly from the scanner for the reconstruction task. The target is derived from the fast inverse Fourier transform on the input data. The AFT does not compress the coil channel so that the input and output shapes/sizes are the same. Network performance is evaluated within magnitude images obtained by the Fourier application and coil compression, which are crucial for radiologists in clinical applications. We also recognize some clinical settings that rely on the phase image, so we also provide evaluation on the phase images.

For the MRI accelerated reconstruction experiment, we first trained AFT, CUNet (DFTNet-I) and DFTNet-K models. Then, we replace the DFT with AFT in the DFTNet to train the AFTNet (I and K) models. At last, we concatenate the pretrained K model with I model to train the KI model. To be more specific, we train an AFT-only network to see if, without convolution layers, the AFT can remove artifacts and enhance quality. Then a network with the AFT followed by the CUNet is trained to simulate a typical deep learning workflow where conventional numerical methods are used to preprocess the image and CNNs are utilized to map the input domain to the target domain. We also evaluate the network with CUNet first implemented directly in the k-space domain. Given that each position in k-space contains the information of the whole image, CNNs implemented in k-space can leverage the complete information of all spaces, even if they have a fixed field of view. Finally, a CUNet-AFT-CUNet structure is evaluated with the first CUNet that extracts k-space domain features and the second CUNet extracting image domain features.

We also extended \ours to 1D scenario for spectral reconstruction to evaluate its generality. In such case, we use the AFTNet-I model and train it from scratch.

\subsection{Experimental data}\label{sec:data}

In this study, two datasets were used: fastMRI dataset~\cite{zbontar2018fastmri}) and Big GABA dataset~\cite{mikkelsen2017big}. The proposed methods were trained on these datasets separately.

The fastMRI dataset contains fully sampled brain MRIs obtained on 3 and 1.5 Tesla magnets. We selected 4-channel axial T1 and T2-weighted scans from the raw fastMRI dataset. We applied a data filtering strategy by counting the number of images with similar shapes and field strength, and selecting the groups that have more than 100 samples. This process resulted in 671 T2-weighted images at 1.5T, 157 T2-weighted images at 3T, and 115 T1-weighted images at 3T. All selected images are acquired on Siemens scanner. A total of 943 scans were used with 794, 99, and 50 each for the training, validation, and test set. The detailed scan parameters are listed in \Cref{tab:a1}. Data preprocessing includes normalization to the maximum intensity value of one in the image domain and cropping to $640 \times 320$. For experiment on accelerated reconstruction, the fully sampled k-space data was undersampled by applying a mask in the phase-encoding direction. We use the acceleration rate (or acceleration factor) to denote the level of scan time reduced for the undersampled k-space data, which is defined as the ratio of the amount of k-space data required for a fully sampled image to the amount collected in an undersampled k-space data~\cite{deshmane2012parallel}. The sampling ratio, SR, is also used to denote the information retained in the undersampled k-space data, which is defined as the inverse of the acceleration rate. An equispaced mask with approximate acceleration matching is used to undersample the k-space data. The fraction of low-frequency columns to be retained for acceleration rates 2x, 4x, and 8x is 16\%, 8\%, and 4\%, respectively.

The\textit{ in vivo} MRS dataset used in this study is from the BIG GABA (\cite{mikkelsen2017big}) repository and contains GABA-edited MEGA-PRESS data acquired on 3T Philips scanners from multiple sites. The voxel was placed in the medial parietal lobe of healthy subjects aged 18 to 36 years. The data consist of 101 pairs of edit-ON/-OFF spectra, each with 160 ON averages and OFF averages. The \ours was trained with an input size of 2048. The ground truth of the ON/OFF spectra is derived by taking the average over 160 transients. We denote the ground truth as noiseless signals. For training, we randomly sampled transients from each subject to reconstruct the averaged signal. By varying the number of transients in computing the average sampled acquisitions, we can generate signals with different levels of signal-to-noise ratio (SNR). We use the term reduction rate (R) to refer to the ratio of the total number of transients and the number of transients used in calculating the average. Being able to reconstruct accurate denoised signals at high R has implications for reducing acquisition time. From a total number of 101 scans in the dataset, 80, 10, and 11 scans were assigned for the training, validation, and test set, respectively. All spectra were normalized to the maximum value of its magnitude. In denoised MRS reconstruction, we use the reduction rate, R, to denote the noise level, which is defined as the ratio of the number of total repetitions (160 for this study) to the number of repetitions retained for a noisy input. We generate noisy free induction decay (FIDs) with 5 reduction rates of 10, 20, 40, 80, and 160.

\subsection{Measurement of Reconstruction Quality}

Three metrics were adopted for the quantitative evaluation of image quality compared to ground truth: structural similarity (SSIM)~\cite{wang2003multiscale}, peak signal-to-noise ratio (PSNR), and normalized root mean squared error (NRMSE). For the quality measurement of the 1D spectra, three other metrics were used: Pearson's correlation coefficient (PCC), Spearman's rank correlation coefficient (SCC) and goodness fitting coefficient (GFC)~\cite{romero1997linear}. The GFC is introduced to evaluate the goodness of the mathematical reconstruction with a value ranging from 0 to 1, where 1 indicates a perfect reconstruction. If $\hat{y}_i$ is the predicted value of the $i$-th sample and $y_i$ is the corresponding true value, then the GFC estimated over $n_{\text{samples}}$ is defined as
\begin{equation}
\text{GFC}(y, \hat{y}) = \frac{|\sum_{i=0}^{n_\text{samples}-1} y_i \hat{y}_i|}{|\sum_{i=0}^{n_\text{samples}-1} y_i^2|^{1/2} |\sum_{i=0}^{n_\text{samples}-1} \hat{y}_i^2|^{1/2}}.
\end{equation}

\subsection{Implementation details}

We construct a batch size of 1 and optimize the network using the ADAM~\cite{kingma2014adam} optimizer. We use a cosine annealing learning rate scheduler~\cite{DBLP:conf/iclr/LoshchilovH17}. The initial learning rate is $10^{-3}$ and the final learning rate is $10^{-5}$. We set the negative slope of LeakReLU to $0.1$. All methods are trained for 50 epochs. All experiments were performed with PyTorch 2.2.2 and a Quadro RTX 6000 GPU. 

For experiment on the MRI accelerated reconstruction, we compared \ours with zero-filling (DFT), numerical method GRAPPA~\cite{griswold2002generalized}, real-valued model KIKI-net~\cite{eo2018kiki} and diffusion model score-MRI~\cite{chung2022score}. For experiment on the MRS denoised reconstruction, we compared \ours with Gaussian line broadening (GLB).

\section{Results}

\subsection{Comparison study}

\begin{figure}[t]
\centering
\includegraphics[width=\linewidth]{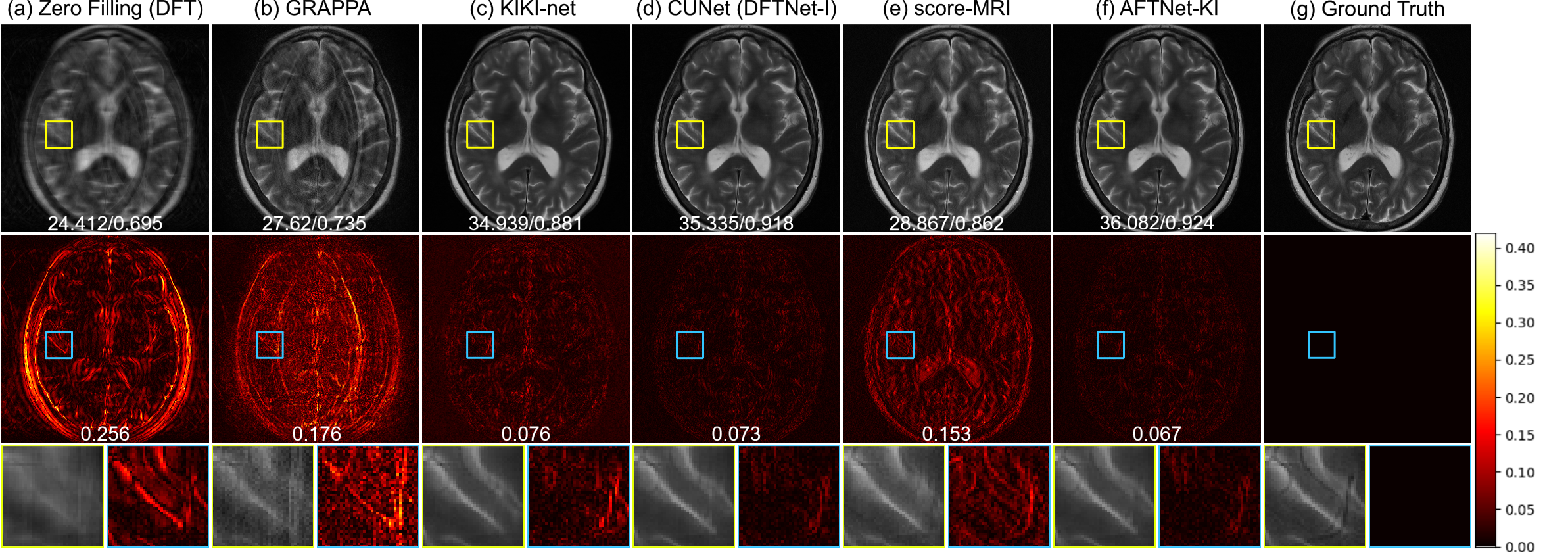}
\caption{Comparison between baseline methods and \ours. Here we present the qualitative results f a T2w image on 3T} under 4x acceleration rate, where a 1D 4x equal-spaced sampling mask with 8\% of low-frequency columns retained is implemented to undersample the input k-space data. Residual map shows the difference magnitude against ground truth (in Hot colormap). Yellow and blue boxes show the zoomed-in version of the indicated area. White numbers in the magnitude images indicate PSNR (dB) and SSIM. White numbers in the residual maps indicate NRMSE.
\label{fig:comparison}
\end{figure}

The effectiveness of \ours was evaluated on the test set of the human normal-field MRI dataset through a comparative study as shown in \Cref{fig:comparison}, where the up row shows the visualization of the baseline models, including conventional Fourier transform-based networks and score-MRI. The bottom row shows the proposed networks and ground truth derived from the fully sampled k-space data. Both the residual maps against the ground truth and the zoomed-in areas are presented. The residual maps are normalized to the maximum value of $0.5$ such that they are comparable across all models. We first compare the AFT-only network with the results after DFT. It clearly shows that the light-trainable AFT block not only approximates the DFT but also facilitates artifact removal with the presence of signal non-ideality. Comparing the results of the AFT-only network and score-MRI, which solves the image reconstruction inverse problem based on score-based generative models, we demonstrate that even with such a simple structure, AFT can achieve similar performance in terms of PSNR and NRMSE. The score-MRI was originally trained on the knee dataset and we fine-tuned it on our dataset. Qualitatively, \ours outperformed score-MRI, as can be seen from the difference magnitude map and the zoomed-in area in \Cref{fig:comparison} by comparing the results of \ours and score-MRI. Quantitative metrics demonstrate the superior performance of \ours over score-MRI and KIKI-net, as shown in \Cref{tab:comparison}, including SSIM, PSNR, and NRMSE in all acceleration rates. The statistical t-test between the score-MRI and \ours metrics also shows the superior accelerated reconstruction of the proposed method with a p-value below 0.0001. Furthermore, the AFT block can serve as a better replacement for DFT by comparing the results of DFTNet and \ours, where all \ours shows better performance than DFTNet, as can be seen from the residual maps. It should be pointed out that \ours-KI significantly outperforms DFTNet-KI, indicating that AFT can be implemented to connect two networks while preserving the capability of both networks.

\subsection{Ablation study}

Different \ours structures, as mentioned in \Cref{sec:framework}, were compared in datasets with different field strengths and different modalities to verify the stability and generality of the \ours. In addition, the effectiveness of the front-end/back-end convolutional networks is also evaluated in this section. To validate the robustness of \ours to k-space artifacts, these proposed \ours structures were compared on image reconstruction and accelerated reconstruction as described in \Cref{sec:data}. Furthermore, the extended \ours was compared with numerical methods using 1-dimensional MRS FID data on the denoised reconstruction.

\begin{figure}[t]
\centering
\includegraphics[width=.5\linewidth]{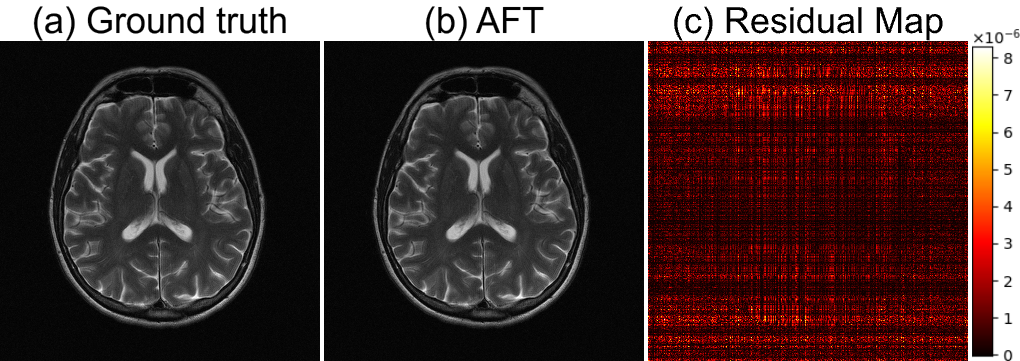}
\caption{Fully sampled MRI reconstruction results. Here we present the qualitative results of a T2w image on 1.5T. (a) Ground truth, (b) proposed method, (c) difference magnitude of (a) and (b) (in Hot colormap).}
\label{fig:recon}
\end{figure}

First, we show the results of human 1.5/3T MRI reconstruction using fully sampled fastMRI k-space data in \Cref{fig:recon}. All the images shown here and in the following sections are cropped so that the anti-aliasing placed outside the field of view (FOV) in phase-encoding directions is removed. The ground truth image is derived by applying the conventional Fourier transformation to the k-space data. It can be seen that the ground truth image obtained from FT is identical to the AFT prediction, which human observers can not distinguish. The results adhere to the mathematical description we discussed in \Cref{sec:cnet}. The residual map (pixel-wise difference between the ground truth image and the AFT prediction) shows that no brain structural information is presented. The grid-like remaining error is mainly caused by precision loss during floating-point calculation in matrix multiplication.

\begin{figure}[t]
\centering
\includegraphics[width=.85\linewidth]{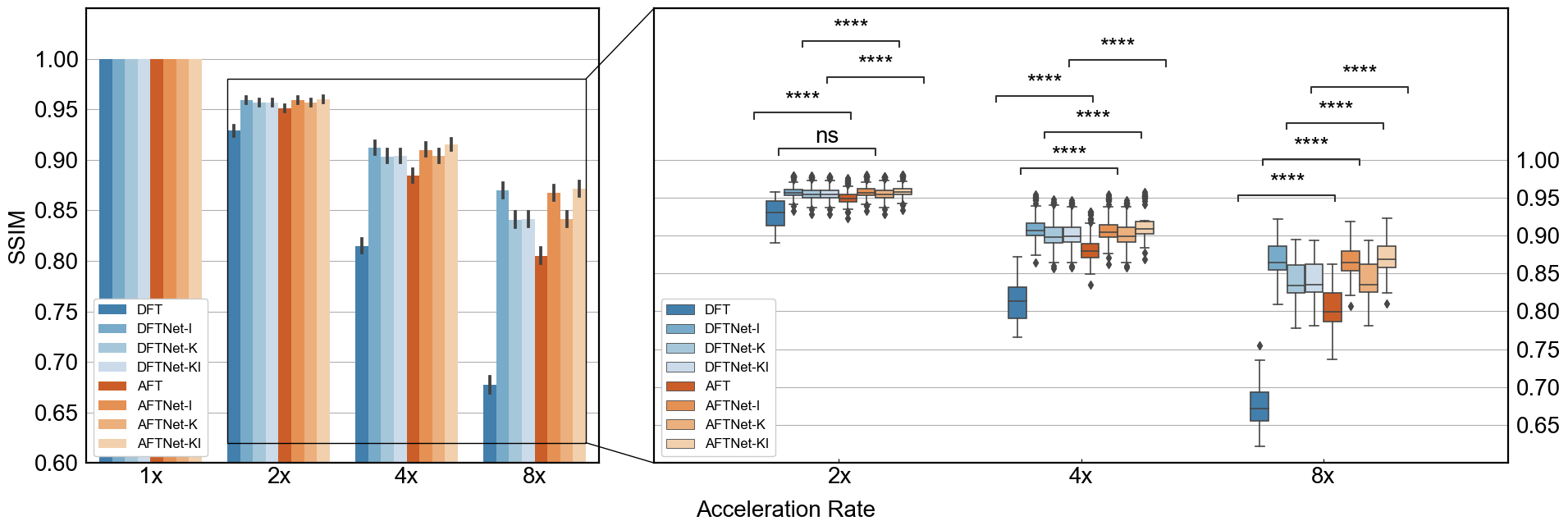}
\caption{Results of human 1.5/3T MRI accelerated reconstruction by comparing \ours (K Model, I Model, and KI Model) and with their DFT counterparts in terms of SSIM for acceleration rates 1x, 2x, 4x, and 8x. p-values indicate results from two-sided t-tests for paired samples between the DFT-based model and AFT-based model. (ns: $p > 0.05$, *: $p \le 0.05$, **: $p \le 0.01$, ***: $p \le 0.001$, ****: $p \le 0.0001$) }
\label{fig:aftnet-vs-dftnet}
\end{figure}

\begin{figure}[t]
\centering
\includegraphics[width=\linewidth]{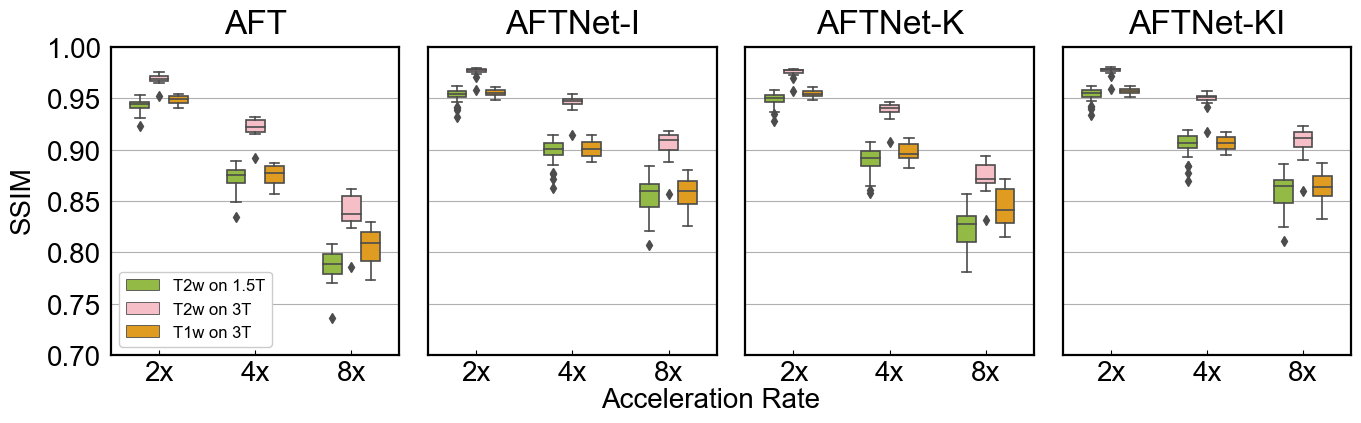}
\caption{Results of human 1.5/3T MRI accelerated reconstruction by comparing \ours (I Model, K Model, and KI Model) in terms of SSIM on different acquisition types and system field strength for acceleration rates 2x, 4x, and 8x.}
\label{fig:aftnet-modality-field}
\end{figure}

In \Cref{fig:comparison}, we show the results of MRI accelerated reconstruction using under-sampled fastMRI k-space data. In the first row, we see the reconstructions from the 1D 4x equal space sampling, in which 8\% of low-frequency columns are retained (see \Cref{fig:a0}). Here, we compare different \ours structures with the zero-filling method. \ours-KI performs outstanding reconstruction, where less structural difference can be seen from the residual map in the second row. The third row shows the zoomed-in areas of both the images and residual maps. \ours-I produces a more blurry reconstruction which loses the structural details. Reconstruction through \ours-K induces foggy artifacts, which are reflected in terms of SSIM. \Cref{fig:aftnet-vs-dftnet} shows the results of the accelerated reconstruction comparing AFT and \ours (I, K, and KI) with their counterparts in terms of SSIM in acceleration rates of 2x, 4x and 8x. The performance of DFT drops linearly as the acceleration rate increases, while the \ours methods are more robust to the acceleration scale. The t-test between AFT-based and DFT-based structures indicates that \ours significantly outperforms all DFTNet structures except the I model in the 2x acceleration rate. The results of \ours on different acquisition types and system field strength are shown in \Cref{fig:aftnet-modality-field}. The difference of performance is mainly caused by the image quality between the 3T and 1.5T images and the inrisint contrast difference between T2w and T1w images.

A comprehensive comparison of quantitative metrics on the test set is provided in \Cref{tab:comparison} and \Cref{tab:aftnet-modality-field} for accelerated reconstruction. In addition to the metrics evaluated on the magnitude images, we provide the evaluation on the phase images in \Cref{tab:phase}. \ours-KI outperforms other \ours structures on all the different acceleration rates, and both \ours-KI and \ours-I perform significantly better than other \ours-K structures. \Cref{tab:aftnet-modality-field} shows detailed quantitative metrics of the human 1.5/3T MRI accelerated reconstruction results grouped by image contrast and system field strength. It is worth mentioning that although \ours-K does not outperform other \ours structures in the accelerated reconstruction task, it demonstrates the ability to learn in a sparse frequency domain and its sparse representations with a complex-valued convolutional network.

\subsection{Generality of \ours}\label{sec:mrs}

\begin{figure}[!htbp]
\centering
\includegraphics[width=\linewidth]{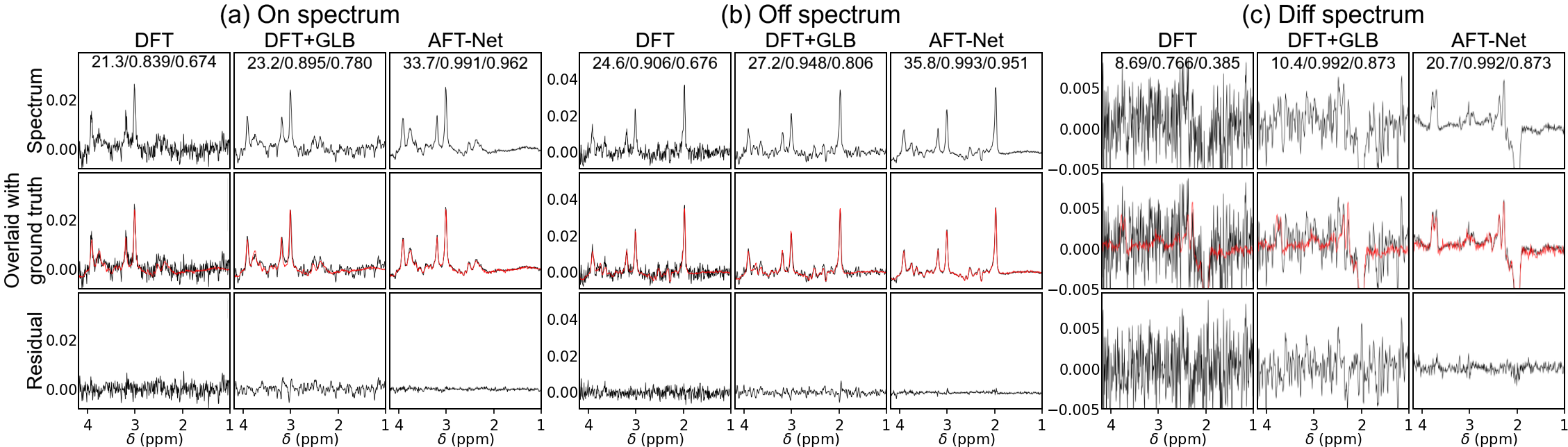}
\caption{Human 3T MRS denoised reconstruction results. The acceleration rate is 80 for each spectrum. (a) Reconstruction results for the ON spectrum, (b) reconstruction results for the OFF spectrum, and (c) results for the DIFF spectrum derived from (a) and (b). 1st row: reconstructed spectra, 2nd row: reconstructed spectra overlaid with ground truth (in red line), 3rd row: Difference of reconstructed spectra against ground truth. Black numbers in the upper center location indicate PSNR (db), PCC, and SCC, respectively.}
\label{fig:mrs-one-sample}
\end{figure}

\begin{figure}[!htbp]
\centering
\includegraphics[width=.65\linewidth]{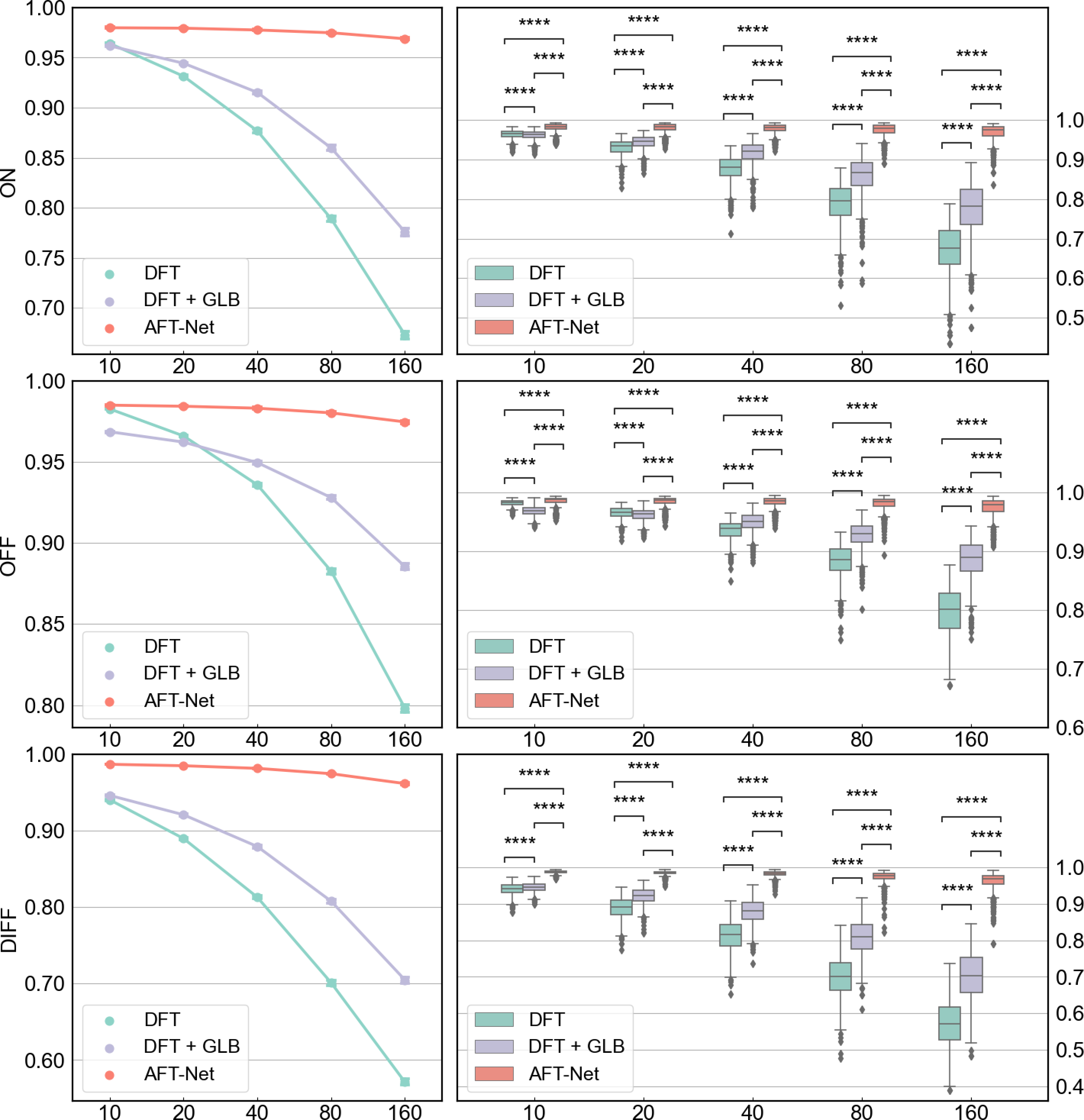}
\caption{Results of human 3T MRS denoised reconstruction by comparing \ours (I Model), DFT, and DFT with Gaussian line broadening in terms of GFC. p-values indicate results from two-sided t-tests for paired samples. (ns: $p > 0.05$, *: $p \le 0.05$, **: $p \le 0.01$, ***: $p \le 0.001$, ****: $p \le 0.0001$)}
\label{fig:mrs-all-samples}
\end{figure}

The results of the proposed \ours approach and traditional DFT method with and without GLB are illustrated in \Cref{fig:mrs-one-sample}. The first row shows the reconstructed spectra from the methods being compared. The second row shows the reconstructed spectra overlaid with the ground truth. The third row shows the difference between the reconstructed spectra and the ground truth. The reconstructed spectra illustrated here represent a reduction rate of 80, where only two out of a total of 160 transients were used to construct the spectrum. The \ours approach outperforms the other methods in all reduction rates listed in the table. We used the GFC to measure the similarity between the reconstructed spectra and the ground truth, as shown in \Cref{tab:mrs}. Although the GFC decreases as the reduction rate increases, the \ours method is able to maintain a high performance even at very high reduction rates. For example, the difference in GFC between the OFF spectra under a reduction rate of 10 and 160 is very small, around 0.9798 and 0.9688, respectively.

\section{Discussion}

In this study, a unified MR image reconstruction framework is proposed, which is composed of two main components: artificial Fourier transform block and complex-valued residual attention UNet. The AFT block is used to approximate conventional DFT, which is demonstrated in \Cref{tab:comparison} and shown in \Cref{sec:AFT}. The front-end/back-end convolutional layers are used to extract features at different levels in the k-space/image domains and play different roles in various tasks. As shown in \Cref{tab:comparison}, both front- and back-end convolutional layers show superior accelerated reconstruction performance under all sampling ratios compared to single front-end/back-end convolutional layers. This is potentially because the undersampling is performed in k-space where the artifacts are separated from the non-artifact. While in the image domain, it is converted to an aliasing overlapped over the whole image. The artifact removal task can be recast as an image inpainting problem in the k-space domain which can be done more easily by the front-end convolutional layers. However, all structures with front-end-only convolutional layers show lower performance, indicating that the sparse representation of k-space data makes it harder for a convolutional network to extract noise information in the low-frequency areas and back-end convolutional layers are necessary to achieve the optimal performance.

Domain-transform manifold learning has been introduced for years, and several deep learning frameworks were developed based on this idea. The first model, AUTOMAP~\cite{zhu2018image}, proposed the simple FC-Conv structure which can only be applied to images with a small matrix size due to its redundant FC layers. DOTA-MRI~\cite{eo2020accelerating} extended AUTOMAP to Conv-FC-Conv structure. However, it did not extend the model structure and implement CVNNs. The \ours we proposed in this study solves the problem mentioned above through a modular-designed AFT block. We also demonstrated that the extended \ours can also be applied to 1D data in \Cref{sec:mrs}. In addition, previous works define the loss in the magnitude image, while we calculate the loss in the complex-valued image domain, which preserves the relations between the real and imaginary parts. The phase is then derived from the output of \ours, which is essential for several phase-based applications, such as flow quantification and fat–water separation.

CVNNs, especially complex-valued convolutional networks~\cite{cole2021analysis,sikka2021cu}, have been studied for MRI reconstruction, but have mainly focused on simple tasks or only applied them to the image domain. We investigate the different impacts of complex-valued convolutional networks on the k-space and image domain and extend the application to accelerated reconstruction and denoised reconstruction, which are more clinically important. We also incorporate domain-manifold learning by adding domain transform blocks that determine the mapping between the k-space and the image domain instead of the conventional discrete Fourier transform. It is more robust to noise and signal non-ideality due to imperfect acquisition. We also extend the application of complex-valued convolutional networks to 1D MRS denoised reconstruction, which has not been studied in previous work. Our results indicate that \ours is able to effectively decrease the contribution of the noise in the FIDs while preserving the quantifiable signals in the reconstructed spectra.

One remaining methodological limitation is that the FC layers used by \ours narrow the application to datasets with various image matrix sizes. Although the convolutional layers are not sensitive to the image matrix sizes and cropping/padding can be applied to match the desired sizes, the features of FC layers need to be selected carefully which requires further investigation. Another parameter that needs to be taken into account is the coil number. In this study, we selected especially 4-channel MRI data for convenience of data preprocessing, while deep learning-based coil combinations could be incorporated into the framework in future work. Furthermore, diffusion models are shown to be a powerful tool for image reconstruction across body regions and coil numbers~\cite{chung2022score}. However, the score-MRI we compared in this study does not demonstrate superior performance compared with \ours and the inference stage time is long. This is potential because the backbone of the score-MRI is still a real-valued UNet structure and the relation between the real and imaginary part is not considered during the calculation of the score function. For future works, the \ours could be further extended by leveraging diffusion-based models with complex-valued convolutional networks as the backbone and careful optimization to reduce the inference time.

\section{Conclusion}

In conclusion, we propose AFT, a novel artificial Fourier transform framework that determines the mapping between k-space and image domain as conventional DFT while having the ability to be fine-tuned/optimized with further training. The flexibility of AFT allows it to be easily incorporated into any existing deep learning network as learnable or static blocks. We then utilized AFT to design our \ours, which implements complex-valued UNet to extract features in the k-space and/or image domain. We aim to combine reconstruction and acceleration/denoising tasks into a unified network that simultaneously enhances the image quality by removing artifacts directly from the k-space and/or image domain. The proposed methods are evaluated on datasets with additional artifacts, different contrasts, and different modalities. Our \ours achieves competitive results compared with other methods and is found to be robust to noise differences. An extensive study on various system fields, various modalities, and various tasks demonstrates the effectiveness and generality of \ours. Our current claims are supported by quantitative and objective visual analyses. In future works, neuroradiologist evaluations could be incorporated to further strengthen the qualitative assessment.

\bibliographystyle{elsarticle-num} 
\bibliography{ref}

\begin{thebibliography}{10}
\expandafter\ifx\csname url\endcsname\relax
  \def\url#1{\texttt{#1}}\fi
\expandafter\ifx\csname urlprefix\endcsname\relax\def\urlprefix{URL }\fi
\expandafter\ifx\csname href\endcsname\relax
  \def\href#1#2{#2} \def\path#1{#1}\fi

\bibitem{georgiou1992complex}
G.~Georgiou, C.~Koutsougeras, Complex domain backpropagation, IEEE Transactions on Circuits and Systems II: Analog and Digital Signal Processing (1992).

\bibitem{guberman2016complex}
N.~Guberman, On complex valued convolutional neural networks, arXiv preprint arXiv:1602.09046 (2016).

\bibitem{trabelsi2018deep}
C.~Trabelsi, O.~Bilaniuk, Y.~Zhang, D.~Serdyuk, S.~Subramanian, J.~F. Santos, S.~Mehri, N.~Rostamzadeh, Y.~Bengio, C.~J. Pal, Deep complex networks, International Conference on Learning Representations (2018).

\bibitem{johnson1928thermal}
J.~B. Johnson, Thermal agitation of electricity in conductors, Physical review (1928).

\bibitem{nyquist1928thermal}
H.~Nyquist, Thermal agitation of electric charge in conductors, Physical review (1928).

\bibitem{hansen2015image}
M.~S. Hansen, P.~Kellman, Image reconstruction: an overview for clinicians, Journal of Magnetic Resonance Imaging (2015).

\bibitem{fessler2010model}
J.~A. Fessler, Model-based image reconstruction for mri, IEEE signal processing magazine (2010).

\bibitem{de2016machine}
M.~De~Bruijne, Machine learning approaches in medical image analysis: From detection to diagnosis (2016).

\bibitem{shen2017deep}
D.~Shen, G.~Wu, H.-I. Suk, Deep learning in medical image analysis, Annual review of biomedical engineering (2017).

\bibitem{zhu2018image}
B.~Zhu, J.~Z. Liu, S.~F. Cauley, B.~R. Rosen, M.~S. Rosen, Image reconstruction by domain-transform manifold learning, Nature (2018).

\bibitem{lopez2021spiking}
J.~L{\'o}pez-Randulfe, T.~Duswald, Z.~Bing, A.~Knoll, Spiking neural network for fourier transform and object detection for automotive radar, Frontiers in Neurorobotics (2021).

\bibitem{eo2020accelerating}
T.~Eo, H.~Shin, Y.~Jun, T.~Kim, D.~Hwang, Accelerating cartesian mri by domain-transform manifold learning in phase-encoding direction, Medical Image Analysis (2020).

\bibitem{chung2022score}
H.~Chung, J.~C. Ye, Score-based diffusion models for accelerated mri, Medical image analysis (2022).

\bibitem{zhao2023swingan}
X.~Zhao, T.~Yang, B.~Li, X.~Zhang, Swingan: A dual-domain swin transformer-based generative adversarial network for mri reconstruction, Computers in Biology and Medicine (2023).

\bibitem{yan2023dc}
Y.~Yan, T.~Yang, X.~Zhao, C.~Jiao, A.~Yang, J.~Miao, Dc-siamnet: Deep contrastive siamese network for self-supervised mri reconstruction, Computers in Biology and Medicine (2023).

\bibitem{xu2020learning}
K.~Xu, M.~Qin, F.~Sun, Y.~Wang, Y.-K. Chen, F.~Ren, Learning in the frequency domain, Proceedings of the IEEE/CVF conference on computer vision and pattern recognition (2020).

\bibitem{hirose2012complex}
A.~Hirose, Complex-valued neural networks, Wiley Online Library, 2012.

\bibitem{hirose2012generalization}
A.~Hirose, S.~Yoshida, Generalization characteristics of complex-valued feedforward neural networks in relation to signal coherence, IEEE Transactions on Neural Networks and learning systems (2012).

\bibitem{tygert2016mathematical}
M.~Tygert, J.~Bruna, S.~Chintala, Y.~LeCun, S.~Piantino, A.~Szlam, A mathematical motivation for complex-valued convolutional networks, Neural computation (2016).

\bibitem{ronneberger2015u}
O.~Ronneberger, P.~Fischer, T.~Brox, U-net: Convolutional networks for biomedical image segmentation, Medical Image Computing and Computer-Assisted Intervention--MICCAI 2015: 18th International Conference, Munich, Germany, October 5-9, 2015, Proceedings, Part III 18 (2015).

\bibitem{sikka2021cu}
D.~Sikka, N.~Igra, S.~Gjerwold-Sellec, C.~Gao, E.~Wu, J.~Guo, Cu-net: A completely complex u-net for mr k-space signal processing, ISMRM (International Society of Magnetic Resonance Imaging) Virtual Conference \& Exhibition, 2021 (2021).

\bibitem{cole2021analysis}
E.~Cole, J.~Cheng, J.~Pauly, S.~Vasanawala, Analysis of deep complex-valued convolutional neural networks for mri reconstruction and phase-focused applications, Magnetic resonance in medicine (2021).

\bibitem{ioffe2015batch}
S.~Ioffe, C.~Szegedy, Batch normalization: Accelerating deep network training by reducing internal covariate shift, International conference on machine learning (2015).

\bibitem{wu2018group}
Y.~Wu, K.~He, Group normalization, Proceedings of the European conference on computer vision (ECCV) (2018).

\bibitem{ba2016layer}
J.~L. Ba, J.~R. Kiros, G.~E. Hinton, Layer normalization, arXiv preprint arXiv:1607.06450 (2016).

\bibitem{roemer1990nmr}
P.~B. Roemer, W.~A. Edelstein, C.~E. Hayes, S.~P. Souza, O.~M. Mueller, The nmr phased array, Magnetic resonance in medicine (1990).

\bibitem{ma2024magnetic}
D.~J. Ma, Y.~Yang, N.~Harguindeguy, Y.~Tian, S.~A. Small, F.~Liu, D.~L. Rothman, J.~Guo, Magnetic resonance spectroscopy spectral registration using deep learning, Journal of Magnetic Resonance Imaging (2024).

\bibitem{wu2023deep}
Z.~Wu, W.~Liao, C.~Yan, M.~Zhao, G.~Liu, N.~Ma, X.~Li, Deep learning based mri reconstruction with transformer, Computer Methods and Programs in Biomedicine (2023).

\bibitem{zbontar2018fastmri}
J.~Zbontar, F.~Knoll, A.~Sriram, T.~Murrell, Z.~Huang, M.~J. Muckley, A.~Defazio, R.~Stern, P.~Johnson, M.~Bruno, et~al., fastmri: An open dataset and benchmarks for accelerated mri, arXiv preprint arXiv:1811.08839 (2018).

\bibitem{mikkelsen2017big}
M.~Mikkelsen, P.~B. Barker, P.~K. Bhattacharyya, M.~K. Brix, P.~F. Buur, K.~M. Cecil, K.~L. Chan, D.~Y.-T. Chen, A.~R. Craven, K.~Cuypers, et~al., Big gaba: Edited mr spectroscopy at 24 research sites, Neuroimage (2017).

\bibitem{deshmane2012parallel}
A.~Deshmane, V.~Gulani, M.~A. Griswold, N.~Seiberlich, Parallel mr imaging, Journal of Magnetic Resonance Imaging (2012).

\bibitem{wang2003multiscale}
Z.~Wang, E.~P. Simoncelli, A.~C. Bovik, Multiscale structural similarity for image quality assessment, The Thrity-Seventh Asilomar Conference on Signals, Systems \& Computers, 2003 (2003).

\bibitem{romero1997linear}
J.~Romero, A.~Garc{\i}a-Beltr{\'a}n, J.~Hern{\'a}ndez-Andr{\'e}s, Linear bases for representation of natural and artificial illuminants, JOSA A (1997).

\bibitem{kingma2014adam}
D.~P. Kingma, J.~Ba, Adam: A method for stochastic optimization, arXiv preprint arXiv:1412.6980 (2014).

\bibitem{DBLP:conf/iclr/LoshchilovH17}
I.~Loshchilov, F.~Hutter, {SGDR:} stochastic gradient descent with warm restarts, 5th International Conference on Learning Representations, {ICLR} 2017, Toulon, France, April 24-26, 2017, Conference Track Proceedings (2017).

\bibitem{griswold2002generalized}
M.~A. Griswold, P.~M. Jakob, R.~M. Heidemann, M.~Nittka, V.~Jellus, J.~Wang, B.~Kiefer, A.~Haase, Generalized autocalibrating partially parallel acquisitions (grappa), Magnetic Resonance in Medicine: An Official Journal of the International Society for Magnetic Resonance in Medicine (2002).

\bibitem{eo2018kiki}
T.~Eo, Y.~Jun, T.~Kim, J.~Jang, H.-J. Lee, D.~Hwang, Kiki-net: cross-domain convolutional neural networks for reconstructing undersampled magnetic resonance images, Magnetic resonance in medicine (2018).

\end{thebibliography}

\section*{Tables}

\begin{table}[!htbp]
\centering
\resizebox{\textwidth}{!}{
\begin{tabular}{llllllllllll} 
\hline
\multirow{2}{*}{Acc.} & \multirow{2}{*}{Metrics} & \multicolumn{5}{l}{Baseline} &  & \multicolumn{4}{l}{Ours} \\ 
\cline{3-7}\cline{9-12}
 &  & DFT & Grappa & KIKI-net & score-MRI & DFTNet-I &  & AFT & \ours-K & \ours-I & \ours-KI \\ 
\hline
\multirow{3}{*}{1x} & SSIM & \textbf{-} & 1.000 & 0.998 & \textbf{1.000} & \textbf{1.000} &  & \textbf{1.000} & \textbf{1.000} & \textbf{1.000} & \textbf{1.000} \\
 & PSNR & \textbf{-} & \textbf{161.08} & 46.8 & 63.9 & 153.3 &  & 153.3 & 153.3 & 153.3 & 153.3 \\
 & NRMSE & \textbf{-} & 0.000 & 0.024 & 0.003 & \textbf{0.000} &  & \textbf{0.000} & \textbf{0.000} & \textbf{0.000} & \textbf{0.000} \\ 
\hline
\multirow{3}{*}{2x} & SSIM & 0.929 & 0.928 & 0.948 & 0.939 & 0.959 &  & 0.951 & 0.957 & 0.959 & \textbf{0.960}$^****$ \\
 & PSNR & 33.6 & 35.27 & 39.2 & 38.5 & 39.5 &  & 37.0 & 38.9 & 39.6 & \textbf{39.8}$^****$ \\
 & NRMSE & 0.105 & 0.088 & 0.06 & 0.052 & 0.053 &  & 0.070 & 0.057 & 0.052 & \textbf{0.051}$^****$ \\ 
\hline
\multirow{3}{*}{4x} & SSIM & 0.815 & 0.835 & 0.870 & 0.874 & 0.912 &  & 0.885 & 0.904 & 0.912 & \textbf{0.915}$^****$ \\
 & PSNR & 27.3 & 29.16 & 34.7 & 32.8 & 35.2 &  & 32.0 & 33.7 & 35.3 & \textbf{35.7}$^****$ \\
 & NRMSE & 0.214 & 0.173 & 0.092 & 0.099 & 0.086 &  & 0.125 & 0.103 & 0.085 & \textbf{0.082}$^****$ \\ 
\hline
\multirow{3}{*}{8x} & SSIM & 0.677 & 0.744 & 0.783 & 0.787 & 0.869 &  & 0.805 & 0.841 & 0.869 & \textbf{0.872}$^****$ \\
 & PSNR & 23.8 & 25.29 & 31.4 & 27.5 & 31.3 &  & 27.6 & 29.4 & 31.3 & \textbf{31.5}$^****$ \\
 & NRMSE & 0.32 & 0.271 & 0.134 & 0.189 & 0.135 &  & 0.208 & 0.171 & 0.137 & \textbf{0.132}$^****$ \\
\hline
\end{tabular}
}
\caption{Quantitative metrics of human 1.5/3T MRI accelerated reconstruction for magnitude images. Numbers are presented as mean value ± standard deviation. Numbers in boldface indicate the best metric out of all the methods. Acc.: Acceleration Rate. p-values indicate results from two-sided t-tests for paired samples between best and second-best models. (ns: $p > 0.05$, *: $p \le 0.05$, **: $p \le 0.01$, ***: $p \le 0.001$, ****: $p \le 0.0001$)}
\label{tab:comparison}
\end{table}

\begin{table}[!htbp]
\centering
\resizebox{\textwidth}{!}{%
\begin{tabular}{lllllllllll}
\hline
\multirow{2}{*}{Acc.} & \multirow{2}{*}{Metrics} & \multicolumn{4}{l}{Baseline} &  & \multicolumn{4}{l}{Ours} \\ \cline{3-6} \cline{8-11} 
 &  & DFT & DFTNet-K & DFTNet-I & DFTNet-KI &  & AFT & \ours-K & \ours-I & \ours-KI \\ \hline
\multirow{3}{*}{1x} & SSIM & \textbf{-} & \textbf{1.000} & \textbf{1.000} & \textbf{1.000} & \textbf{} & \textbf{1.000} & \textbf{1.000} & \textbf{1.000} & \textbf{1.000} \\
 & PSNR & \textbf{-} & \textbf{126.36} & \textbf{126.36} & \textbf{126.36} &  & \textbf{126.36} & \textbf{126.36} & \textbf{126.36} & \textbf{126.36} \\
 & NRMSE & \textbf{-} & \textbf{0.000} & \textbf{0.000} & \textbf{0.000} &  & \textbf{0.000} & \textbf{0.000} & \textbf{0.000} & \textbf{0.000} \\ \hline
\multirow{3}{*}{2x} & SSIM & 0.687 & 0.730 & 0.746 & 0.749 &  & 0.717 & 0.730 & 0.747 & \textbf{0.752}$^{****}$ \\
 & PSNR & 10.50 & 11.00 & 11.24 & 11.28 &  & 10.81 & 11.00 & 11.26 & \textbf{11.32}$^{****}$ \\
 & NRMSE & 0.671 & 0.634 & 0.617 & 0.614 &  & 0.647 & 0.635 & 0.615 & \textbf{0.611}$^{****}$ \\ \hline
\multirow{3}{*}{4x} & SSIM & 0.585 & 0.610 & 0.666 & 0.669 &  & 0.633 & 0.610 & 0.669 & \textbf{0.672}$^{****}$ \\
 & PSNR & 9.31 & 9.59 & 10.03 & 10.06 &  & 9.65 & 9.59 & 10.06 & \textbf{10.10}$^{****}$ \\
 & NRMSE & 0.766 & 0.743 & 0.708 & 0.705 &  & 0.738 & 0.743 & 0.705 & \textbf{0.702}$^{****}$ \\ \hline
\multirow{3}{*}{8x} & SSIM & 0.561 & 0.561 & 0.618 & 0.621 &  & 0.590 & 0.559 & 0.620 & \textbf{0.623}$^{****}$ \\
 & PSNR & 8.84 & 8.87 & 9.33 & 9.36 &  & 9.07 & 8.87 & 9.34 & \textbf{9.40}$^{***}$ \\
 & NRMSE & 0.807 & 0.805 & 0.766 & 0.763 &  & 0.787 & 0.805 & 0.765 & \textbf{0.760}$^{***}$ \\ \hline
\end{tabular}%
}
\caption{Quantitative metrics of human 1.5/3T MRI accelerated reconstruction for phase images. Numbers are presented as mean value ± standard deviation. Numbers in boldface indicate the best metric out of all the methods. Acc.: Acceleration Rate. p-values indicate results from two-sided t-tests for paired samples between best and second-best models. (ns: $p > 0.05$, *: $p \le 0.05$, **: $p \le 0.01$, ***: $p \le 0.001$, ****: $p \le 0.0001$)}
\label{tab:phase}
\end{table}

\begin{table}[!htbp]
\begin{subtable}[h]{\textwidth}
\centering
\caption{T2w images on 1.5T}
\resizebox{\textwidth}{!}{%
\begin{tabular}{lllllll}
\hline
\multirow{2}{*}{Acc.} & \multirow{2}{*}{Metrics} & \multirow{2}{*}{DFT} & Ours &  &  &  \\ \cline{4-7} 
 &  &  & AFT & \ours-K & \ours-I & \ours-KI \\ \hline
\multirow{3}{*}{1x} & SSIM & \textbf{-} & \textbf{1.000 ± 0.000} & \textbf{1.000 ± 0.000} & \textbf{1.000 ± 0.000} & \textbf{1.000 ± 0.000} \\
 & PSNR & \textbf{-} & \textbf{154.0 ± 1.3} & \textbf{154.0 ± 1.3} & \textbf{154.0 ± 1.3} & \textbf{154.0 ± 1.3} \\
 & NRMSE & \textbf{-} & \textbf{0.000 ± 0.000} & \textbf{0.000 ± 0.000} & \textbf{0.000 ± 0.000} & \textbf{0.000 ± 0.000} \\ \hline
\multirow{3}{*}{2x} & SSIM & 0.911 ± 0.011 & 0.943 ± 0.008 & 0.948 ± 0.008 & 0.952 ± 0.008 & \textbf{0.953   ± 0.007} \\
 & PSNR & 32.5 ± 1.1 & 36.5 ± 1.2 & 38.0 ± 1.3 & 39.1 ± 1.2 & \textbf{39.2   ± 1.2} \\
 & NRMSE & 0.127 ± 0.008 & 0.080 ± 0.004 & 0.068 ± 0.005 & 0.060 ± 0.004 & \textbf{0.059   ± 0.005} \\ \hline
\multirow{3}{*}{4x} & SSIM & 0.792 ± 0.014 & 0.871 ± 0.013 & 0.890 ± 0.014 & 0.897 ± 0.014 & \textbf{0.903 ± 0.013} \\
 & PSNR & 26.7 ± 1.0 & 31.7 ± 1.4 & 32.8 ± 1.3 & 35.0 ± 1.2 & \textbf{35.3   ± 1.2} \\
 & NRMSE & 0.248 ± 0.015 & 0.141 ± 0.006 & 0.124 ± 0.007 & 0.095 ± 0.007 & \textbf{0.093 ± 0.007} \\ \hline
\multirow{3}{*}{8x} & SSIM & 0.662 ± 0.023 & 0.788 ± 0.016 & 0.823 ± 0.020 & 0.855 ± 0.019 & \textbf{0.859   ± 0.018} \\
 & PSNR & 23.6 ± 1.1 & 27.0 ± 1.4 & 28.4 ± 1.5 & \textbf{30.9 ± 1.3} & \textbf{30.9   ± 1.3} \\
 & NRMSE & 0.358 ± 0.022 & 0.240 ± 0.011 & 0.204 ± 0.010 & \textbf{0.153 ± 0.008} & \textbf{0.153 ± 0.009} \\ \hline
\end{tabular}
}
\end{subtable}
\begin{subtable}[h]{\textwidth}
\centering
\caption{T2w images on 3T}
\resizebox{\textwidth}{!}{%
\begin{tabular}{lllllll}
\hline
\multirow{2}{*}{Acc.} & \multirow{2}{*}{Metrics} & \multirow{2}{*}{DFT} & Ours &  &  &  \\ \cline{4-7} 
 &  &  & AFT & \ours-K & \ours-I & \ours-KI \\ \hline
\multirow{3}{*}{1x} & SSIM & \textbf{-} & \textbf{1.000 ± 0.000} & \textbf{1.000 ± 0.000} & \textbf{1.000 ± 0.000} & \textbf{1.000   ± 0.000} \\
 & PSNR & \textbf{-} & \textbf{152.3 ± 1.2} & \textbf{152.3 ± 1.2} & \textbf{152.3 ± 1.2} & \textbf{152.3 ± 1.2} \\
 & NRMSE & \textbf{-} & \textbf{0.000 ± 0.000} & \textbf{0.000 ± 0.000} & \textbf{0.000 ± 0.000} & \textbf{0.000   ± 0.000} \\ \hline
\multirow{3}{*}{2x} & SSIM & 0.948 ± 0.007 & 0.968 ± 0.006 & 0.975 ± 0.006 & \textbf{0.976 ± 0.006} & \textbf{0.976 ± 0.006} \\
 & PSNR & 33.1 ± 1.1 & 36.6 ± 1.3 & 39.4 ± 1.5 & 40.2 ± 1.5 & \textbf{40.4 ± 1.6} \\
 & NRMSE & 0.093 ± 0.009 & 0.063 ± 0.006 & 0.046 ± 0.006 & 0.042 ± 0.006 & \textbf{0.041 ± 0.006} \\ \hline
\multirow{3}{*}{4x} & SSIM & 0.841 ± 0.021 & 0.921 ± 0.011 & 0.937 ± 0.011 & 0.945 ± 0.010 & \textbf{0.948   ± 0.010} \\
 & PSNR & 26.7 ± 1.2 & 31.1 ± 1.3 & 33.3 ± 1.4 & 35.2 ± 1.4 & \textbf{35.5 ± 1.5} \\
 & NRMSE & 0.195 ± 0.017 & 0.117 ± 0.007 & 0.091 ± 0.006 & 0.074 ± 0.008 & \textbf{0.071   ± 0.008} \\ \hline
\multirow{3}{*}{8x} & SSIM & 0.694 ± 0.031 & 0.838 ± 0.021 & 0.872 ± 0.017 & 0.903 ± 0.017 & \textbf{0.906 ± 0.017} \\
 & PSNR & 23.1 ± 1.3 & 26.6 ± 1.3 & 28.2 ± 1.3 & 30.5 ± 1.2 & \textbf{30.7 ± 1.3} \\
 & NRMSE & 0.295 ± 0.022 & 0.198 ± 0.011 & 0.164 ± 0.009 & 0.126 ± 0.010 & \textbf{0.124   ± 0.010} \\ \hline
\end{tabular}
}
\end{subtable}
\begin{subtable}[h]{\textwidth}
\centering
\caption{T1w images on 3T}
\resizebox{\textwidth}{!}{%
\begin{tabular}{lllllll}
\hline
\multirow{2}{*}{Acc.} & \multirow{2}{*}{Metrics} & \multirow{2}{*}{DFT} & Ours &  &  &  \\ \cline{4-7} 
 &  &  & AFT & \ours-K & \ours-I & \ours-KI \\ \hline
\multirow{3}{*}{1x} & SSIM & \textbf{-} & \textbf{1.000 ± 0.000} & \textbf{1.000 ± 0.000} & \textbf{1.000 ± 0.000} & \textbf{1.000   ± 0.000} \\
 & PSNR & \textbf{-} & \textbf{153.0 ± 0.8} & \textbf{153.0 ± 0.8} & \textbf{153.0 ± 0.8} & \textbf{153.0 ± 0.8} \\
 & NRMSE & \textbf{-} & \textbf{0.000 ± 0.000} & \textbf{0.000 ± 0.000} & \textbf{0.000 ± 0.000} & \textbf{0.000   ± 0.000} \\ \hline
\multirow{3}{*}{2x} & SSIM & 0.941 ± 0.006 & 0.949 ± 0.004 & 0.955 ± 0.004 & 0.956 ± 0.004 & \textbf{0.957 ± 0.003} \\
 & PSNR & 35.8 ± 0.7 & 38.3 ± 0.7 & 40.0 ± 0.8 & 40.1 ± 0.9 & \textbf{40.4 ± 0.8} \\
 & NRMSE & 0.079 ± 0.008 & 0.059 ± 0.004 & 0.048 ± 0.003 & 0.048 ± 0.004 & \textbf{0.046 ± 0.003} \\ \hline
\multirow{3}{*}{4x} & SSIM & 0.830 ± 0.014 & 0.875 ± 0.010 & 0.898 ± 0.009 & 0.901 ± 0.008 & \textbf{0.906   ± 0.008} \\
 & PSNR & 28.9 ± 0.7 & 33.2 ± 0.8 & 35.7 ± 0.9 & 35.9 ± 1.1 & \textbf{36.4 ± 1.0} \\
 & NRMSE & 0.174 ± 0.020 & 0.106 ± 0.008 & 0.080 ± 0.006 & 0.077 ± 0.009 & \textbf{0.073   ± 0.007} \\ \hline
\multirow{3}{*}{8x} & SSIM & 0.687 ± 0.024 & 0.805 ± 0.019 & 0.844 ± 0.019 & 0.858 ± 0.016 & \textbf{0.863 ± 0.015} \\
 & PSNR & 24.7 ± 0.9 & 29.5 ± 0.8 & 32.1 ± 0.9 & 33.0 ± 1.3 & \textbf{33.3 ± 1.2} \\
 & NRMSE & 0.281 ± 0.027 & 0.162 ± 0.013 & 0.121 ± 0.009 & 0.109 ± 0.015 & \textbf{0.104   ± 0.013} \\ \hline
\end{tabular}
}
\end{subtable}
\caption{Quantitative metrics of human 1.5/3T MRI accelerated reconstruction. Numbers are presented as mean value ± standard deviation. Numbers in boldface indicate the best metric out of all the methods.}\label{tab:aftnet-modality-field}
\end{table}

\begin{table}[!htbp]
\centering
\begin{tabular}{lllll}
\hline
R & Spectrum & DFT & DFT+GLB & \ours \\ \hline
10 & ON & 0.9827 ± 0.0047 & 0.9686 ± 0.0086 & \textbf{0.9850 ± 0.0085} \\
 & OFF & 0.9641 ± 0.0104 & 0.9617 ± 0.0108 & \textbf{0.9798 ± 0.0124} \\
 & DIFF & 0.9403 ± 0.0164 & 0.9461 ± 0.0126 & \textbf{0.9868 ± 0.0037} \\
20 & ON & 0.9660 ± 0.0090 & 0.9622 ± 0.0111 & \textbf{0.9843 ± 0.0092} \\
 & OFF & 0.9314 ± 0.0192 & 0.9443 ± 0.0170 & \textbf{0.9794 ± 0.0127} \\
 & DIFF & 0.8897 ± 0.0283 & 0.9208 ± 0.0206 & \textbf{0.9849 ± 0.0055} \\
40 & ON & 0.9359 ± 0.0162 & 0.9496 ± 0.0155 & \textbf{0.9831 ± 0.0098} \\
 & OFF & 0.8768 ± 0.0318 & 0.9152 ± 0.0281 & \textbf{0.9776 ± 0.0139} \\
 & DIFF & 0.8129 ± 0.0418 & 0.8792 ± 0.0325 & \textbf{0.9815 ± 0.0078} \\
80 & ON & 0.8826 ± 0.0280 & 0.9280 ± 0.0214 & \textbf{0.9803 ± 0.0120} \\
 & OFF & 0.7890 ± 0.0486 & 0.8598 ± 0.0452 & \textbf{0.9748 ± 0.0160} \\
 & DIFF & 0.7010 ± 0.0566 & 0.8077 ± 0.0480 & \textbf{0.9745 ± 0.0154} \\
160 & ON & 0.7981 ± 0.0403 & 0.8854 ± 0.0325 & \textbf{0.9747 ± 0.0160} \\
 & OFF & 0.6730 ± 0.0619 & 0.7759 ± 0.0638 & \textbf{0.9688 ± 0.0200} \\
 & DIFF & 0.5710 ± 0.0654 & 0.7047 ± 0.0662 & \textbf{0.9616 ± 0.0245} \\ \hline
\end{tabular}
\caption{GFC metric of human 3T MRS denoised reconstruction. Numbers are presented as mean value ± standard deviation. Numbers in boldface indicate the best metric out of all the methods.}
\label{tab:mrs}
\end{table}

\clearpage
\newpage

\appendix
\setcounter{table}{0}
\setcounter{figure}{0}
\section{}

\begin{table}[!htbp]
\centering
\resizebox{\columnwidth}{!}{%
\begin{tabular}{cccccccccc}
\hline
\begin{tabular}[c]{@{}c@{}}System\\ Model\end{tabular} & \begin{tabular}[c]{@{}c@{}}Field \\ Strength (T)\end{tabular} & \begin{tabular}[c]{@{}c@{}}Institution\\ Name\end{tabular} & \begin{tabular}[c]{@{}c@{}}Protocol\\ Name\end{tabular} & \begin{tabular}[c]{@{}c@{}}Encoded\\ Matrix Size\end{tabular} & \begin{tabular}[c]{@{}c@{}}Field of View\\ (mm)\end{tabular} & \begin{tabular}[c]{@{}c@{}}TR\\ (ms)\end{tabular} & \begin{tabular}[c]{@{}c@{}}TE\\ (ms)\end{tabular} & \begin{tabular}[c]{@{}c@{}}Sequence\\ Type\end{tabular} & Count \\ \hline
Avanto & 1.494 & NYU & AX & 640 x 272 x 1 & 440 x 186.9725 x 7.5 & 5120 & 103 & TurboSpinEcho & 1 \\
Avanto & 1.494 & NYU & AX & 640 x 320 x 1 & 440 x 220 x 7.5 & 4000 & 107 & TurboSpinEcho & 1 \\
Avanto & 1.494 & NYU & AX & 640 x 320 x 1 & 440 x 220 x 7.5 & 5120 & 103 & TurboSpinEcho & 373 \\
Avanto & 1.494 & NYU & AX & 640 x 320 x 1 & 440 x 220 x 7.5 & 5120 & 107 & TurboSpinEcho & 295 \\
Avanto & 1.494 & NYU Medical Center & AX T2\_FBB & 640 x 320 x 1 & 440 x 220 x 7.5 & 5120 & 103 & TurboSpinEcho & 1 \\
TrioTim & 2.8936 & NYU & AX & 768 x 308 x 1 & 440 x 176.5392 x 7.5 & 6000 & 107 & TurboSpinEcho & 1 \\
TrioTim & 2.8936 & NYU & AX & 768 x 350 x 1 & 440 x 200.4446 x 7.5 & 6000 & 107 & TurboSpinEcho & 1 \\
TrioTim & 2.8936 & NYU & AX & 768 x 392 x 1 & 440 x 224.62 x 7.5 & 6000 & 107 & TurboSpinEcho & 141 \\
TrioTim & 2.8936 & NYU & AX & 768 x 392 x 1 & 460 x 234.83 x 7.5 & 6000 & 107 & TurboSpinEcho & 14 \\
TrioTim & 2.8936 & NYU Clinical Cancer Ctr & AX T1 PRE\_FBB & 640 x 260 x 1 & 440 x 178.75 x 5 & 264 & 2.88 & Flash & 1 \\
TrioTim & 2.8936 & NYU Clinical Cancer Ctr & AX T1 PRE\_FBB & 640 x 320 x 1 & 440 x 220 x 5 & 250 & 2.88 & Flash & 1 \\
TrioTim & 2.8936 & NYU Clinical Cancer Ctr & AX T1 PRE\_FBB & 640 x 320 x 1 & 440 x 220 x 5 & 264 & 2.88 & Flash & 106 \\
TrioTim & 2.8936 & NYU Clinical Cancer Ctr & AX T1 PRE\_FBB & 640 x 320 x 1 & 460 x 230 x 5 & 264 & 2.88 & Flash & 5 \\
TrioTim & 2.8936 & NYU Clinical Cancer Ctr & AX T1 PRE\_FBB & 640 x 320 x 1 & 480 x 240 x 5 & 264 & 2.88 & Flash & 2 \\ \hline
\end{tabular}
}
\caption{Imaging parameters. The name AX indicates T2w imageing protocol.}
\label{tab:a1}
\end{table}

\begin{figure}[!htbp]
\centering
\includegraphics[width=.4\textwidth]{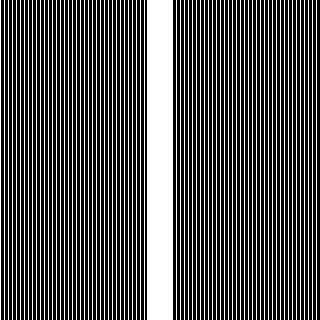}
\caption{1D 4x equal-spaced sampling mask with 8\% of low-frequency columns retained. While space indicates the signal retained and black space indicates the signal masked out.}
\label{fig:a0}
\end{figure}

\end{document}